\documentstyle[prb,aps]{revtex}

\newcommand{\be}{\begin{equation}}              
\newcommand{\ee}{\end{equation}}                
\newcommand{\bea}{\begin{eqnarray}}             
\newcommand{\eea}{\end{eqnarray}}
\newcommand{\nn}{\nonumber}
\newcommand{\bm}[1]{\mbox{\boldmath${#1}$}}
\newcommand{\R}{\bm{R}}

\newcommand{\grad}{\bm{\nabla}}
\tighten                                

\begin{document}


\title{Effective Vortex Dynamics in Superfluid Systems}
\author{C. Wexler and D. J. Thouless}
\address{Department of Physics, Box 351560, 
         University of Washington, 
         Seattle, WA 98195-1560}
\date{December 1996}
\maketitle

\begin{abstract}
An alternative approach to the derivation of the force on a vortex 
based in an adiabatic approximation in the action of the
superfluid system is developed. Assuming that the vortex motion
is relatively slow compared with the characteristic times
involved in the microscopic degrees of freedom, the effect of the
superfluid and its excitations is reduced to a gauge potential, and
the associated transverse force. 
The vortex velocity part of the transverse force is found in terms of
the thermal expectation of the angular momentum of the fluid around
the vortex. The excitations countercirculate the vortex reducing the
effective density to that of the superfluid part only.
Non-adiabatic contributions appear in
this formalism as a non-Abelian gauge potential connecting
different microscopic states, in particular the renormalization of the
vortex mass is given by the lowest order diagonal correction to the
adiabatic theory, and the long wavelength contributions found to 
depend only on the static structure factor for the fluid. 
\end{abstract}

\pacs{47.37.+q,67.40.Vs,67.57.Fg,76.60.Ge}

\section*{Introduction}

Quantized vortices have been an essential part in the theory of
superfluids since Onsager first stated the idea of quantized
circulation in the late 40's \cite{vort_intro,feynman55}. 
In fact, vortices in general have been an important part of classical 
fluid mechanics for decades\cite{saffman92}, in particular in the 
study of turbulent regimes.
The Magnus force , or Kutta-Joukowski hydrodynamic lift, 
$\bm{F}_M = \rho \; \bm{\kappa} \times (\bm{v}_{V} - 
\bm{v}_{\text{fluid}})$ 
is well known from classical hydrodynamics and occurs whenever an 
object with circulation $\kappa$ around it moves through a
fluid\cite{lamb32}, an important application being the lift force 
on an airplane wing. The coupling to the fluid is also known to 
produce a {\em hydrodynamic mass}.

The problem of obtaining effective dynamics for adiabatic and
non-adiabatic motion of vortices is highly controversial. In particular, the 
expression for the Magnus force at finite temperatures is still far
from clear, and today (more than three decades after the first
measurements by Vinen\cite{vinen61}), different 
expressions for the Magnus force can be found in the 
literature\cite{barenghi83,qvih2,volovik95,demircan95,tan96,sonin96}. 
The confusion arises, in part, from different 
interpretations on the role played
by excitations being scattered asymmetrically at the vortex, leading
to a  transverse force proportional to the normal fluid density
$\rho_n$ and either the relative velocity $(\bm{v}_{\text{n}} - 
\bm{v}_{V})$ or $(\bm{v}_{\text{n}} - \bm{v}_{\text{s}})$,
namely the Iordanskii force\cite{iordanskii66}. 
The magnitude of this term must be calculated
and, moreover, it is not clear whether it should be added to the
Magnus force written above, with the coefficient $\rho$, or to a
similar expression involving the superfluid part $\rho_s$ only.
This paper analyzes the situation for neutral, homogeneous
superfluids,  while the case of charged superfluids and
non translation invariant systems will be published 
elsewhere\cite{geller}. 

For fermion superfluids, Volovik\cite{volovik93} and
Stone\cite{stone96} have identified additional contributions from
localized quasiparticles inside the vortex core. Thouless, Ao and 
Niu\cite{tan96}(TAN) show that no such a contribution exists for
translation invariant systems, as seems to be observed\cite{zieve93}
for the $B$ phase of $\rm^3$He. 

Another point of controversy is the effective mass.
Coupling of the vortex motion and the superfluid circulating around it
renormalize the mass, and it has been argued by some authors that this
renormalized mass is logarithmically divergent with the system
size\cite{duan}, while others estimate it to be finite\cite{muirhead} or even
zero\cite{volovik72}. Recent studies by us and Demircan 
{\em et al.}\cite{demircan96} show that the vortex mass is
logarithmically dependent on the frequency, cutting off the size
dependence for large systems.

In section \ref{sec:eff_act} we develop a formalism for the 
{\em effective action} of
a vortex moving relative to the fluid. Following TAN, this is 
accomplished by the introduction of a short
range repulsive potential for the particles, strong enough to pin the
vortex (a simple example would be a macroscopic solid wire as done in
Vinen's experiment\cite{vinen61}). We follow closely the adiabatic
approximation as done by Moody, Shapere and Wilczek\cite{moody}, 
and arrive to a simple {\em exact} expression for the effective action
involving the vortex coordinates. Expansion of this action yields 
the {\em vortex velocity part of the Magnus force} (VVPMF) 
to first order in the vortex velocity (sec. \ref{sec:vvpmf}). 
This force depends on the {\em total circulation} around the vortex, 
which is calculated in section \ref{sec:tot_circ}. The excitations
countercirculate the vortex in a way that is equivalent to stating
that only the superfluid fraction has a non-vanishing circulation as
stated in TAN. The VVPMF is given by $\rho_s \bm{\kappa}_s \times
\bm{v}_{V}$.

The normal and superfluid velocity parts of the Magnus force (NVPMF, 
SVPMF) require extra work, and will be dealt with in a future
publication. 

Section \ref{sec:mass} deals with the second order terms, which 
renormalize the vortex mass. This mass renormalization depends only on
the static structure factors for the fluid. We recover a
logarithmically divergent mass in the zero-frequency limit.

\section{Adiabatic Effective Action for a Vortex}
\label{sec:eff_act}

Following reference TAN\cite{tan96}, we control the motion of a
vortex by means of a pinning potential, consisting in a short-range
repulsive potential for the particles that form the fluid. At this
stage we only require this interaction to be a pure potential
$V(\R,\eta)$ without any velocity dependent part, where $\R$ is 
the vortex coordinate and $\eta \equiv \{\bm{r}_i\}$ is the set of all
particle coordinates.
We follow  Moody, Shapere and Wilczek\cite{moody} rather closely, 
avoiding some sign inconsistencies along the way.
We can immediately write the Lagrangian of the system as

\be
L = L_{\text{bare}} + l_{\text{fluid}}(\R,\eta) ,
\ee

\noindent
where  $L_{\text{bare}}$ is the Lagrangian of the bare pinning center
and $l_{\text{fluid}}$ incorporates the full 
Lagrangian of the fluid and the repulsive potential
$V(\R,\eta)$. Notice that we have imposed no condition on the form of
the Lagrangians, except for the type of interaction {\em between} the
pinning center and the fluid particles.

The full time-evolution kernel of the system can now be written 
in terms of Feynman path integrals over all possible
configurations\cite{feynman64}:

\be
U(\R_f,\eta_f,t_f;\R_i,\eta_i,t_i) \equiv
\langle  \R_f,\eta_f,t_f | \R_i,\eta_i,t_i \rangle
=       \int_{\R_i}^{\R_f} \!\!
        \int_{\eta_i}^{\eta_f} {\cal D}[\R]\:{\cal D}[\eta]\: \exp 
        \left\{ \frac{i}{\hbar} \; \int_{t_i}^{t_f} [ 
                L_{\text{bare}} + l_{\text{fluid}}(\R,\eta) ] dt
        \right\} .
\ee

For a stationary vortex at position $\R$ we can also, in principle, obtain
the exact eigenstates of the fluid subsystem

\be
h(\R) \phi_n (\R) = \epsilon_n (\R) \phi_n (\R) ,
\ee

\noindent
and for any particular value of $\R$, any wavefunction for the fast
system can be expanded in this complete basis 

\be
\psi = \sum_n \phi_n (\R) F_n (\R) .
\ee

We will be interested in finding a kernel $U_{mn} (\R_f,
t_f;\R_i,t_i)$ that links the final weights $F_m(\R_f,t_f)$ with the 
initial state's $F_n(\R_i,t_i)$:

\be
F_m(\R_f,t_f) =  
        \sum_n U_{mn} (\R_f, t_f;\R_i,t_i) \; F_n(\R_i,t_i) .
\ee

This kernel has all the information that we
need, and is diagonal in the stationary case ${\bf\dot{\R}}=0$. Most
important is the fact that, for non-degenerate cases, off-diagonal
contributions are small. This new kernel is given by

\bea 
U_{mn} (\R_f, t_f;\R_i,t_i) && \equiv \int \!\!\! \int d{\eta_f}\: d{\eta_i} 
        \; \langle \phi_m t_f | {\eta_f} \rangle        \;
        U(\R_f,\eta_f,t_f;\R_i,\eta_i,t_i) \;
        \langle {\eta_i} | \phi_n t_i \rangle  \nn \\
&& = \int_{\R_i}^{\R_f} {\cal D}[\R] \;
        \; e^{ \frac{i}{\hbar} \; \int_{t_i}^{t_f} L_{\text{bare}}} 
        \; U_{mn}^{\text{fluid}} ,
\eea
\be
U_{mn}^{\text{fluid}}   \equiv \int \!\!\! \int d{\eta_f} \: d{\eta_i}\;
        \langle \phi_m t_f | {\eta_f} \rangle \:
        \langle {\eta_i} | \phi_n t_i \rangle
        \int_{\eta_i}^{\eta_f} {\cal D}[\eta] \;
        e^{ \frac{i}{\hbar}\; \int_{t_i}^{t_f} l_{\text{fluid}}(R,\eta)  dt} 
= \langle \phi_m t_f | \phi_n t_i \rangle .
\ee

We can calculate this propagator in the usual way by dividing the time
interval $[t_i,t_f]$ into $N$ segments and inserting the identity
operator at each time. Taking $N \rightarrow \infty$ yields

\bea
U_{mn}^{\text{fluid}} && =
\langle \phi_m t_f |
        \sum_{l_1, \dots, l_{N-1}} \prod_{j=1,\dots, N-1}
        \exp \left\{  \frac{i}{\hbar} \; \Delta t \; [ 
                - \epsilon_{l_j} \: \delta_{l_j, l_{j-1}} 
                + i \; \hbar \; {\bf\dot{\R}}(t_j) \cdot 
                \langle \phi_{l_j} t_j | \grad_R \phi_{l_{j-1}} t_j \rangle
                ] \right\} \;
        |\phi_n t_i \rangle   \nn \\
&& = T \exp \left\{   \frac{i}{\hbar} \; \int_{t_i}^{t_f} dt \; 
        [- \epsilon_m \:\delta_{mn}+
        i \; \hbar \; {\bf\dot{\R}}(t) \cdot \langle  \phi_m | 
        \grad_R \phi_n \rangle ]
        \right\} ,
\eea

\noindent
where the last expression is formal way of denoting
an infinite number of time ordered matrix products. 

The time evolution of the pinning center  can now be written in terms
of an effective action

\be
U_{mn} (\R_f, t_f;\R_i,t_i) = 
\int_{\R_i}^{\R_f} {\cal D}[\R] \;
        T \; e^{  \frac{i}{\hbar} \; S_{mn}^{\text{eff}} } ,
\ee

\be
S_{mn}^{\text{eff}} = \int_{t_i}^{t_f} dt \; [
         L_{\text{bare}} - \epsilon_m \delta_{mn} 
                + i \; \hbar \; {\bf\dot{\R}} \cdot 
                     \langle \phi_{m}  | 
                        \grad_R \phi_{n} \rangle ] .
\ee

The results insofar are exact, but not very helpful. In general, the
gauge-like term $i \: \hbar \: {\bf\dot{\R}} \cdot \langle \phi_{m} | \grad_R
\phi_{n} \rangle$ is non-Abelian, and we have to deal with complicated
time ordered products, etc. For {\em non-degenerate} states, however,
it is possible to show that non-diagonal terms vanish faster than any
power of $(\tau \Delta \epsilon/\hbar )^{-1}$, with $\tau$ being a
characteristic time of motion of the pinning center, and $\Delta
\epsilon$ the smallest energy difference of the fluid's
dynamics (see ref. \onlinecite{moody}). 
In most cases it will be possible to greatly
simplify the problem by only dealing with a few degenerate or closely
degenerate states.

For non-degenerate cases, transitions to other states may be neglected
and we can easily calculate the effective action for a given state of
the fluid to any order in the vortex velocity. In particular, the
Magnus force is linear in the velocities, and effective masses
correspond to quadratic terms in the action. Expanding to these
orders, the effective action for the vortex motion is simply given by

\be
\label{eq:eff_action}
S_n = \int_{t_i}^{t_f} dt \; \left[ 
        L_{\text{bare}} - \epsilon_n +  i \; \hbar \; {\bf\dot{\R}}(t_j) \cdot
        \langle \phi_{n}  | \grad_R \phi_{n} \rangle + 
        \frac{1}{2} \sum_{ij} M^{ij}_n \dot{R}_i \dot{R}_j \right] ,
\ee

\be
\label{eq:eff_mass}
M^{ij}_n \equiv 2 \; \hbar^2 \;\sum_{l \neq n} 
        \frac{\langle \partial_i \phi_n | \phi_l \rangle
                \langle  \phi_l | \partial_j \phi_n \rangle}
        {\epsilon_l - \epsilon_n} .
\ee

\section{Vortex Velocity part of the Magnus Force}
\label{sec:vvpmf}

From the effective action (\ref{eq:eff_action}), we can
immediately find the transverse VVPMF for any given quantum state $\phi_n$ 
of the fluid:

\be
\bm{F}_n = i \; \hbar \; {\bf \dot{\R}} \times \left( \grad_R \times 
        \langle \phi_n | \grad_R \phi_n \rangle \right) .
\ee

Identifying $\bm{v}_{V} = {\bf \dot{\R}}$, 
we now perform a statistical averaging over initial states to obtain
the thermally averaged VVPMF

\bea
\bm{F} \times \bm{\hat{z}} 
    &&= - i \; \hbar \; \bm{v}_{V}  \;
        \sum_n \; f_n \; \left[ 
        \frac{\partial}{\partial X} \langle \phi_n | 
                \frac{\partial \phi_n}{\partial Y}  \rangle - 
        \frac{\partial}{\partial Y} \langle \phi_n | 
                \frac{\partial \phi_n}{\partial X}  \rangle \right] \nn \\
    && = - i \; \hbar \; \bm{v}_{V} \;
        \sum_n \; f_n \; \left[  
        \langle \frac{\partial \phi_n}{\partial X} |
                \frac{\partial \phi_n}{\partial Y}  \rangle - 
        \langle \frac{\partial \phi_n}{\partial Y} |
                \frac{\partial \phi_n}{\partial X}  \rangle \right] , 
\eea

\noindent 
where $f_n$ is the probability of occupation of the state $\phi_n$. 
This is the familiar form of the Berry curvature\cite{berry84}. For a
homogeneous system it is possible to substitute $\grad_R = - \sum_i
\grad_i$, and therefore

\bea
{F} / v_V
      &&= - i \; \hbar \;
        \sum_{i,j} \sum_n \: f_n \; \left[
        \langle \frac{\partial \phi_n}{\partial x_i} |
                \frac{\partial \phi_n}{\partial y_j}  \rangle - 
        \langle \frac{\partial \phi_n}{\partial y_i} |
                \frac{\partial \phi_n}{\partial x_j}  \rangle \right] \nn\\
      &&= - i \; \hbar \; \bm{\hat{z}} \; \bm{\cdot } \int d^2r [\grad
                \times \grad' \rho(\bm{r}',\bm{r})]_{r=r'} \;
        - \; i \; \hbar \int d^2r \int d^2r [2 \grad_1 \times {\grad_2}'
                \Gamma(\bm{r_1}',\bm{r_2}';\bm{r_1},\bm{r_2})]_{r=r'} ,
\eea

The last term in the equation above vanishes, since the first line
above is the commutator of the $x$ and $y$ components of the total
momentum which is a one particle operator\cite{tan96}. The integrand
of the first term is equal to $\grad \times (\grad - \grad') 
\rho(\bm{r},\bm{r}')/2$. Application of Stokes' theorem can be used to
write the force per unit length

\be
F / v_V
      = - \frac{i  \hbar}{2} \oint_\Gamma d \bm{l} \bm{\cdot} [
        (\grad - \grad') \rho (\bm{r},\bm{r}')]_{r=r'} .
\ee

This result is exact, and the contour of integration $\Gamma$ can be
taken as far from the vortex as one wishes, and there are no
contributions from the vicinity of the core. 

For a neutral superfluid the integrand is just the momentum density
$\bm{j}$.
At this stage one can decompose the momentum density in a two fluid
picture as $\; \bm{j} \equiv \rho_s \: \bm{v}_s + \rho_n \:
\bm{v}_n$, and the vortex velocity part of the Magnus force will be
given by the superfluid and normal mass densities times the superfluid and
normal {\em circulations}

\be
\label{eq:circulations}
F / v_V = \oint_\Gamma ( \rho_s \bm{v}_s + \rho_n \bm{v}_n )
        \bm{\cdot} d \bm{l} = \rho_s \: \kappa_s + \rho_n \: \kappa_n .
\ee

The circulation of the superfluid $\kappa_s$ is quantized to multiples of
$h/m$, and both $\rho_s$ and $\rho_n$ are well defined
quantities. TAN have argued for the normal fluid to
have no circulation. However, the circulation of a normal fluid is 
something that has to be explicitly calculated\cite{pitaevskii96},
which is done in the following section.

\section{Circulation of the Normal Fluid}
\label{sec:tot_circ}

As stated before, one cannot in general {\em assume} the value of the
circulation of a regular fluid. Aeronautical engineers put up a fair
amount of effort to calculate the circulation around a particular 
wing section. Fortunately our case is considerably simpler. We follow 
the method used by Landau to calculate the normal
density\cite{landau_rho_n} in order to
obtain the circulation of the normal fluid.  
The VVPMF is then completely defined in terms of macroscopic transport
parameters.

Note that the expression for the Magnus force is given in terms of
the circulation, and that it can be defined by any path circling the
vortex.Consider a vortex centered in a cylindrical container and
``average'' over all possible paths that go once around the vortex

\be
\label{eq:force_ang_mom}
F / v_V = \frac{1}{L_z} \int_0^{L_z} dz \; \frac{2}{R^2} \int_0^R r dr \;
        \oint_{\Gamma(z,r)} \bm{j} \bm{\cdot} d \bm{l}  = 
        2 \pi \; \frac{1}{\pi R^2 L_z} \; d^3x \: \bm{\hat{z}} 
        \bm{\cdot} (\bm{r} \times \bm{j}) 
        = 2 \pi \frac{{\cal L}_z}{\cal V} . 
\ee

The Magnus force per unit length is just $2 \pi$ times the average
angular momentum ${\cal L}_z$ per unit volume ${\cal V}$. 
At zero temperature, all the fluid
is superfluid, each particle averages and angular momentum $\hbar$.
At zero temperature (or in the ground-state), the  Magnus force is 
simply given by

\be
F_0 / v_V = 2 \pi \:\frac{\hbar N}{\cal V} = \rho \; \frac{h}{m} .
\ee

Finite temperatures can be analyzed by adding the effect of
excitations. At low temperatures excitations can be dealt in a dilute
gas approximation. Moreover, phonons become predominant and we may
neglect rotons. At this stage one wishes to calculate the expectation
value of the angular momentum of the excitations. The presence of the
vortex couples the excitations to the superfluid velocity field,
Doppler shifting the energy of excitations, thus increasing the
occupation of {\em countercirculating} modes relative to those that
circulate in the direction of the superfluid.

The expectation of the angular momentum of the excitations is 
simply given by thermally populating the different phonon modes
(labeled in cylindrical coordinates by the radial and axial
wave-numbers, and by the azimuthal angular momentum):

\be
\label{eq:Lexc}
{ \cal L}_{\text{excit}} = \sum_{k_r,k_z,m} \hbar \: m \;
        \frac{1}{e^{\hbar \omega/k_B T} - 1} .
\ee

The mode frequencies can be obtained from wave equations derived 
from either a Feynman many-body state\cite{noz-pines90} or a
non-linear Gross-Pitaevskii theory \cite{gross-pit61}. From this
latter point of view, we start with a non-linear Schr\"odinger
equation to describe a weakly interacting Bose gas, and after some
algebra (Appendix \ref{sec:gp2phonon}) get the following wave equation
for phonons in presence of a static vortex:

\be
\label{eq:wave_eq_ph}
c^2 \grad^2 \psi - \ddot{\psi} - \frac{\kappa_s}{\pi r}
        \bm{\hat{\varphi}} \; \bm{\cdot} \grad \dot{\psi} = 0 .
\ee

This last term decreases the frequency of phonons countercirculating the
vortex. In a cylindrical geometry, the eigenstates can be labeled by
the angular momentum, radial and azimuthal wave-vectors. To a good
approximation the normal mode frequencies are given by:


\be
\omega \approx c \sqrt{k_r^2 + k_z^2} + m \; \frac{\kappa_s}{2 \pi} \;
        \langle \frac{1}{r^2} \rangle .
\ee

Using this information in equation (\ref{eq:Lexc}) we get to first
order in the superfluid circulation

\be
{ \cal L}_{\text{excit}} \approx  
                - \frac{\hbar^2 \kappa_s}{2 \pi k_B T}  \;
                \sum_{k_r,k_z,m} \; m^2 \;
                \langle \frac{1}{r^2} \rangle
        \frac{e^{\hbar c k/k_B T}}{(e^{\hbar c k/k_B T} - 1)^2} ,
\ee

\noindent
where now the sum is performed on the normal modes {\em in absence} of
the vortex. We now assume that the typical inter-phonon equilibration 
distance is much smaller than the size of our container, which might be 
rather unrealistic at very low temperatures\cite{khalat65}, but
makes sense for a theoretical infinite volume, infinite
time-scale limit. In this limit, equilibration occurs locally, rather
than at the boundaries of the system, and we may simplify the summation
over modes by a sum over uniformly distributed states in phase space
($\sum_{k_r,k_z,m} \rightarrow \int d^3r \: d^3k \: /(2 \pi)^3$, $m
\rightarrow (\bm{r}\times \bm{k})_z$ and $<1/r^2> \rightarrow 1/r^2$):

\be
{ \cal L}_{\text{excit}} \approx  
                - \frac{\hbar^2 \kappa_s}{2 \pi k_B T}  \;
                (\pi R^2 L_z) \; 
                \int_0^{2 \pi} \! \! \sin^2\varphi_k \: d \varphi_k \;
                \int_0^{\infty} \!\! d k_r \: 
                \int_{-\infty}^{\infty} \!\! d k_z \: k_r^3 \: 
        \frac{e^{\hbar c k/k_B T}}{(e^{\hbar c k/k_B T} - 1)^2} =
        - (\pi R^2 L_z) \; 
                \frac{2 \pi^2}{45} \frac{(k_B T)^4}{\hbar^3 c^5} \;
                \frac{\kappa_s}{2 \pi} ,
\ee

\noindent
which shows the excitations are indeed giving some {\em
countercirculation}  around the
vortex. The total angular momentum is just given by the sum of the
ground state (or zero-temperature) angular momentum and that of the
excitations, which for $\kappa_s = h/m$ yields

\be
{\cal L} = {\cal L}_0 + { \cal L}_{\text{excit}} = 
        \frac{\pi R^2 L_z}{2 \pi} \; \frac{h}{m} \; \left[
        \rho - \frac{2 \pi^2}{45} \frac{(k_B T)^4}{\hbar^3 c^5} 
        \right]  = 
        \frac{\pi R^2 L_z}{2 \pi} \; \frac{h}{m} \; 
        [ \rho - \rho_n ] = 
        \frac{\pi R^2 L_z}{2 \pi} \; \frac{h}{m} \; \rho_s ,
\ee

\noindent
where the last two identities follow from identifying the second term
in the brackets as the normal fluid
density due to phonons\cite{landau_rho_n}. This magnitude of the excitation 
countercirculation is equivalent to the non-circulation of the normal fluid
in the two fluid model. Application of equation
(\ref{eq:force_ang_mom}) gives the correct vortex velocity part of the
Magnus force

\be
\label{eq:VVPMF}
\bm{F}_{\text{vortex}} =  \rho_s \: \frac{h}{m} \; \hat{\bm{z}} \times
              \bm{v}_V .
\ee

Where the result above was calculated with $\bm{v}_s = \bm{v}_n =
0$. Next section deals with the coefficient of the force proportional
to $\bm{v}_n$, that is the NVPMF.

\section{Vortex Mass}
\label{sec:mass}

In this section we calculate the renormalization of the vortex
mass in the static limit. Vortex mass calculations have been performed
starting from the hydrodynamic equations\cite{duan}, and here we
approach the problem from a different (albeit equivalent) point of
view: from the second order correction to the adiabatic approximation
in the effective action (eq. \ref{eq:eff_action} and
\ref{eq:eff_mass}). This second order adiabatic approximation
yields correct results for the long wavelength contribution to the
vortex mass. 

For a superfluid, one can write Feynman's many-body
wavefunction\cite{feynman55,feynman56} for a static vortex at $\R$ as

\be
| \psi \rangle \approx \prod_i e^{i \theta (\bm{r}_i - \R)} \; 
        | \phi_0 \rangle ,
\ee

\noindent 
where $| \phi_0 \rangle$ is the many-body ground state. In this 
simplified case

\be
| \partial_X \psi \rangle = \sum_i \frac{i \; y_i}{x_i^2 + y_i^2} \; 
        | \psi \rangle 
= \int d\bm{r} \frac{i \; y}{x^2 + y^2} \sum_i \delta(\bm{r} - \bm{r_i}) \;
        | \psi \rangle 
= \int d\bm{r} \frac{i \; y}{x^2 + y^2} \rho(\bm{r}) | \psi \rangle ,
\ee

\noindent
where $\rho(\bm{r})$ is the density matrix. This can be Fourier
transformed to read

\be
| \partial_X \psi \rangle = - 4 \pi^2  \: \int \frac{d \bm{k'}}{(2 \pi)^3} 
        \delta(k_z') \frac{k_y'}{k'^2} \rho_{\bm{k'}} \;  | \psi \rangle .
\ee

It is clear that the only
significant couplings in eq. (\ref{eq:eff_mass}) 
will be to states that differ from the ``one
vortex ground state'' by $\rho$, that is the one vortex plus one
phonon states:

\be
| \bm{k} \rangle \approx \rho_{\bm{k}} | \psi \rangle .
\ee

We can readily calculate the overlap

\be
{\langle \bm{k} |  \partial_X \psi \rangle} \;   / \;
        {\langle \bm{k} |\bm{k}\rangle^{1/2}} = 
        - 4 \pi^2  \; \int \frac{d\bm{k'}}{(2 \pi)^3} 
                \; \delta(k_z') \frac{k_y'}{k'^2}
        \langle \psi | \rho_{\bm{k}}^{\dag} \rho_{\bm{k'}} | \psi \rangle \;
        / \; \langle \psi | \rho_{\bm{k}}^{\dag} \rho_{\bm{k}} | 
                \psi \rangle^{1/2}
\ee

Under the approximations used so far the correlation in the last term
can be substituted by the ground state correlation

\be
\langle \psi | \rho_{\bm{k}}^{\dag} \rho_{\bm{k'}} | \psi \rangle \approx
\langle \phi_0 | \rho_{\bm{k}}^{\dag} \rho_{\bm{k'}} | \phi_0 \rangle = 
N \; S(k) \delta_{\bm{k\,k'}} ,
\ee

Where $S(k)$ is the static structure factor\cite{noz-pines90} and, 
in neutral superfluids, is equal to $k/(2 m c)$ in the long wavelength
limit ($c$ is the speed of sound). We are left with

\be
{\langle \bm{k} |  \partial_X \psi \rangle} \; / \;
        {\langle \bm{k} |\bm{k}\rangle^{1/2}} = 
        - 4 \pi^2  \; \frac{\rho}{m} \; \delta(k_z) \; \frac{k_y}{k^2} \;
                \left(\frac{S(k)}{N} \right)^{1/2} .
\ee

Given that $\epsilon(k) \approx k^2/(2 m S(k))$, we can write

\be
M_0^{xy} = M_0^{xy} = 0 ,
\ee

\be
M_0^{xx} = M_0^{yy} =  4 \hbar^2 \rho L_z \; \int d^2k \; 
        \frac{k_y^2}{k^6} S(k)^2 
        = \frac{\pi \hbar^2 \rho L_z}{m^2 c^2} \; \int_{k_{min}}^{k_{max}}
                \frac{dk}{k} 
        = \frac{\pi \hbar^2 \rho L_z}{m^2 c^2} \log \left( \frac{R}{a}
                \right) ,
\ee

Where $R$ and $a$ are large and small distances cutoffs. In the static
limit one obtains the size dependent logarithmic divergence of the
mass as predicted by Duan\cite{duan} from the hydrodynamic equations
of the superfluid. A finite frequency calculation\cite{demircan96} 
will cut $R$, by the phonon wavelength $R \rightarrow \lambda = 
2 \pi c/\omega$. 

A very illustrative fact is that the additional
vortex mass is related to the energy of a static vortex by the Einstein
relation $E_{\text{static}} = M_0 \: c^2$. Where 

\be
E_{\text{static}} = \int d\bm{r} \frac{1}{2}\; \rho \; \bm{v}^2 
                = \frac{\pi \hbar^2 \rho}{m^2} \; \log\left(\frac{R}{a}\right) .
\ee

This is not surprising, and should be expected for any wave-equation 
like dynamics like the ones involved for phonons. 

\section{Conclusions}
\label{sec:conclu}

Starting from an adiabatic expansion for the effective action of a
vortex we have obtained, in a closed and coherent way, the vortex
velocity dependent part of the  
transverse force acting on a rectilinear vortex in a neutral and 
homogeneous superfluid.

In this circumstances, the force that depends on the vortex velocity
can be completely determined by conditions far away from the vortex
core and is proportional to the
superfluid density, which is correct in light of Vinen's
experiment\cite{vinen61}. 

Within the formalism used, the renormalization of the vortex mass
depends only on the form of the static structure factor for the
fluid. The obtained result coincides with similar calculations derived
from hydrodynamic equations.

More research is currently under way to completely determine normal and
superfluid velocities dependent part of the transverse force, and to
elucidate the effects
caused by charge and the loss of translation invariance\cite{geller},
as well as the roles of phonon radiative damping and the longitudinal
part of the Iordanskii force.

\acknowledgements

We wish to thank Boris Spivak, Michael Geller and Lev Pitaevskii for
helpful comments on this problem. 
This work was supported by the NSF grant DMR-9528345.

\hrulefill

\appendix

\section{Wave Equation for Phonons in Presence of a Vortex}
\label{sec:gp2phonon}

Starting from the Gross-Pitaevskii nonlinear Scr\"odinger
equation\cite{noz-pines90}

\be
i \hbar \frac{\partial \Psi}{\partial t} = - \frac{\hbar^2}{2m}
        \grad^2 \Psi + \lambda (|\Psi|^2 - \rho_0) \Psi ,
\ee

\noindent
a Madelung transformation\cite{madelung27} $\Psi = \sqrt{\rho} \,
e^{i S} \;$ leaves us with the continuity equation and 
the Euler-Bernoulli equation\cite{lamb32}:

\be
\frac{\partial \rho}{\partial t} + \grad \bf{\cdot} ( \rho \bf{v} ) = 0 ,
\ee

\be
\frac{\hbar}{m} \dot{S} + \frac{\bm{v}^2}{2} + c^2 
        \frac{\rho - \rho_0}{\rho_0} = 0 .
\ee

\noindent
Where $\bm{v} = \hbar/m \; \grad S$ and we have left out the 
``quantum pressure'' term, which is irrelevant for our purposes. 
After some algebra, separating the static phase due to the vortex and
linearizing, one obtains the following equation of
motion for the phonon part of the phase:


\be
c^2 \grad^2 \psi - \ddot{\psi} - \frac{\kappa_s}{\pi r}
        \bm{\hat{\varphi}} \; \bm{\cdot } \grad \dot{\psi} = 0 .
\ee

This expression is not valid in the immediate vicinity of the vortex,
where the assumptions made are not valid. In general this will only be
important when either dealing with short-wavelength excitations or the
s-wave modes, that can reach these regions.





\references

\bibitem{vort_intro} L. Onsager, Nuovo Cimento Suppl. {\bf 6} 249
        (1949); F. London, {\em Superfluids II}, J. Wiley, New York
        (1954).
\bibitem{feynman55} R. P. Feynman, Prog. Low Temp. Phys. {\bf I},
        North-Holland, chap. 1 (1955).
\bibitem{saffman92} P. G. Saffman, {\em Vortex Dynamics}, 
        Cambridge University Press, 1992; and references therein.
\bibitem{lamb32} Sir H. Lamb, {\em Hydrodynamics}, The University
        Press, Cambridge (1932), G. K. Batchelor {\em An Introduction
        to Fluid Mechanics}, Cambridge University Press (1967).
\bibitem{vinen61} For $\rm^4$He see W.F. Vinen, Proc. Roy. Soc. {\bf A260}, 
        218 (1961); S.C. Whitmore and W. Zimmermann, Phys. Rev. {\bf 166}, 
        181 (1968); P.W. Karn, P.W. Starks and W. Zimmermann, Phys. Rev. B 
        {\bf21}, 1797 (1980). For $\rm^3$He see reference \onlinecite{zieve93}.
\bibitem{barenghi83} C. F. Barenghi, R. J. Donnelly and W. F. Vinen,
        J. Low Temp. Phys. {\bf 52}, 189 (1983). 
\bibitem{qvih2} R. J. Donnelly, {\em Quantized Vortices in Helium II},
                Cambridge University Press, 1991.
\bibitem{volovik95} G. E. Volovik, JETP Lett. {\bf 62}, 65 (1995).
\bibitem{demircan95} E. Demircan, P. Ao and Q. Niu, Phys. Rev. B {\bf 52}, 
        476 (1995).
\bibitem{tan96} D. J. Thouless, P. Ao and Q. Niu, Phys. Rev. Lett. {\bf 76}, 
        3758-61 (1996).
\bibitem{sonin96} E. B. Sonin, cond-matt/960699. 
\bibitem{iordanskii66} S. V. Iordanskii, Sov. Phys. JETP {\bf 22}, 160 (1966).
\bibitem{geller} M. Geller, C. Wexler and D. J. Thouless, to be published.
\bibitem{volovik93} G. E. Volovik, Sov. Phys. JETP {\bf 77}, 435 (1993).
\bibitem{stone96} M. Stone, cond-matt/9605197.
\bibitem{zieve93} R. J. Zieve, Y. M. Mukharsky, J. D. Close, J. C. Davis and 
        R. E. Packard, J. Low Temp. Phys. {\bf 91}, 315 (1993).
\bibitem{duan} J.M. Duan and A. Leggett, Phys. Rev. Lett. {\bf 68},
        1216 (1992);  J.M. Duan, Phys. Rev. B {\bf 48}, 333 (1993); 
        J.M. Duan, Phys. Rev. B {\bf 49}, 12381 (1994). 
\bibitem{muirhead} C. M. Muirhead, W. F. Vinen and R. J. Donnelly,
        Philos. Trans. R. Soc. London {\bf A311}, 433 (1984). 
\bibitem{volovik72} G. E. Volovik, JETP Lett. {\bf 15}, 81 (1972).
\bibitem{demircan96} E. Demircan, P. Ao and Q. Niu, cond-mat/9604010.
\bibitem{moody} J. Moody, A. Shapere and F. Wilczek, in {\em Geometric
        Phases in Physics}, edited by A. Shapere and F. Wilczek,
        World Scientific, Singapore, 160 (1989).
\bibitem{feynman64} R. P. Feynman and E. F. Taylor, {\em Quantum
        Mechanics and Path Integrals}, McGraw-Hill, New York, 1964.
\bibitem{berry84} M. V. Berry, Proc. R. Soc. London A {\bf 392}, 45 (1984).
\bibitem{pitaevskii96} L. Pitaevskii, private communication (1996).
\bibitem{landau_rho_n} L. Landau, J. Phys. {\bf V}, 71 (1941).
\bibitem{noz-pines90} P. Nozi\`eres and D. Pines, {\em The Theory of
        Quantum Liquids} vol. II, Addison-Wesley (1990).
\bibitem{gross-pit61} E. P. Gross, Nuovo Cimento {\bf 20}, 454 (1961).
        L. P. Pitaevskii, Sov. Phys. JETP {\bf 12}, 155 (1961).
\bibitem{khalat65} Khalatnikov, {\em An Introduction to the Theory of 
        Superfluidity}, Benjamin, New York (1965).
\bibitem{feynman56} R. P. Feynman and M. Cohen, Phys. Rev. {\bf 102},
        1189 (1956).
\bibitem{madelung27} E. Madelung, Z. Phys. {\bf 40}, 322 (1927).
\end{document}